# Libtt — A Compiler and Runtime for Tenstorrent Processors

*Philipp Moritz (pcmoritz@relativity.cc) and Sarah Dong (sarah@relativity.cc)*

ABSTRACT

Libtt brings the PyTorch and JAX ecosystem to Tenstorrent hardware. It implements PJRT, the interface through which XLA-based frameworks drive Google's TPUs, so JAX runs on it directly, PyTorch runs through TorchTPU, and software written for TPUs carries over with no changes except replacing custom kernels. Built on Tenstorrent's open-source compiler and runtime, libtt ships the entire stack, from the compiler to the multi-chip runtime, as a single shared library. It contains extensive fixes to the upstream Tenstorrent software to make existing PyTorch and JAX code work out of the box, and extensive new compiler optimizations to make it fast. It passes most of the JAX and TorchTPU test suites and runs SGLang, TorchTitan, and Miles, covering serving, fine-tuning, and reinforcement learning.

## 1. Introduction

Libtt is a shared library that makes a Tenstorrent processor [1] look, to JAX [2] and to PyTorch [3], like a TPU. A supported program that calls `jax.jit` or `torch.compile` or uses eager execution runs on the chip without modifications. The library is loaded through PJRT, the interface by which frameworks built on the XLA compiler talk to a device [4]. It holds everything needed beneath that interface: a compiler from StableHLO [5], the set of tensor operations in which those frameworks hand a program over, to programs for the chip; the run-time system that loads and starts those programs on multiple chips; and the programs' kernels (Figure 1).

The design goal is that existing model code runs unchanged, and that good performance comes out of the box. Rather than asking users to adapt their models to the hardware, libtt places the work of making a model fast in the compiler and the run-time system: the compiler decides how operations are lowered, fused, and mapped onto the chip, while the run-time system manages program loading, data movement, and execution. Because these optimizations operate on StableHLO rather than on any particular model, they are not tied to the models used to develop them. They apply across models whose operations libtt supports.

We demonstrate this design on large language model training and serving. TorchTitan's TPU trainer shards and trains models across multiple chips. Two existing serving engines run on libtt: SGLang-JAX [6], written in JAX, and SGLang-TorchTPU [7], written in PyTorch for Google's TPUs. Neither was written for Tenstorrent hardware. SGLang-JAX gains a Tenstorrent backend that calls libtt's kernels for attention and a few other operations through custom calls, while SGLang-TorchTPU runs without source changes. Together they serve the Qwen3 family [8] and its successors, covering dense models, mixture-of-experts models, and hybrids that combine attention with recurrent layers. Training results are in Section 9; serving results are in Section 10.5 and Section 11. We modeled the software stack on the TPU's because the two architectures share the properties that matter most to the compiler (see Table 1). Both compute on tiles, so most operations require tensors to be stored in tiled layouts in accelerator memory, and both connect chips with high-speed interconnects that favor mesh-style partitioning across devices.

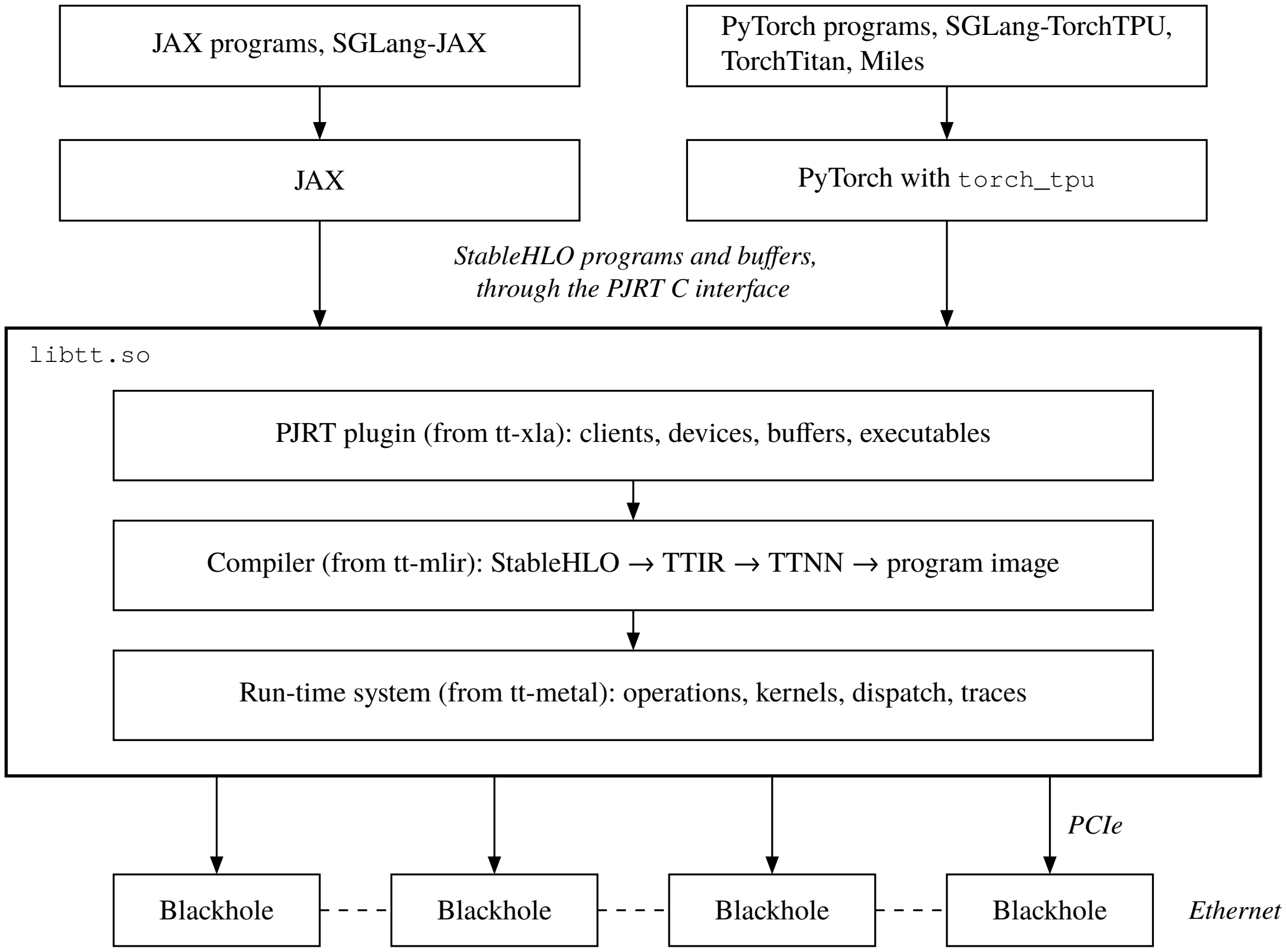

**Figure 1.** The path of a program. Everything inside the large box is one shared library.

Libtt builds on Tenstorrent's PJRT plugin [9], MLIR compiler [10], and tt-metal runtime system [11]. These projects provide the compiler, tensor operations, driver, and kernels, but offer very limited compatibility with existing JAX and PyTorch programs. Libtt addresses these gaps by redesigning parts of the stack, fixing bugs, and adding missing functionality. These changes broaden framework compatibility and integrate the components into a single, capable runtime for both JAX and PyTorch (Section 3, Section 4, Section 5, Section 8, Section 11). Libtt's contributions are:

1. **One drop-in runtime for JAX and PyTorch.** A self-contained PJRT library bundles the compiler, runtime, firmware, and kernel sources, and supports PyTorch through the existing `torch_tpu` backend. Applications use their framework APIs without the need to rewrite them using the Python bindings of Tenstorrent's TTNN library (Section 4, Section 8).
2. **Wide framework coverage.** General indexing kernels and extensions to reductions, sorting, and control flow make a wide range of JAX and PyTorch programs executable. Compatibility is checked against the upstream JAX and `torch_tpu` test suites (Section 5, Section 11) and by running existing training and serving workloads.
3. **Numerical fidelity across the stack.** FP16 and the E4M3FN and E5M2 FP8 formats retain their types through buffers, compilation, and execution; data movement preserves bit patterns and supported matrix products read FP8 operands natively. A vector-unit matrix product provides FP32 arithmetic when highest precision is requested, and integer collectives and cumulative sums preserve exact integer values (Section 5.3, Section 7). In addition, many small numerical precision bugs were fixed.

4. **Multi-chip execution.** Tensor partitioning and collectives support both JAX's single-process model and PyTorch's one-process-per-chip model (Section 7). To enable this, we redesigned the lower-level communication primitives to support the multiple-program, multiple-data execution model often used in PyTorch.
5. **Optimized from compiler to kernels.** Compiler fusions, row-aware kernels, weight preparation, and traces reduce program launches and data movement. We also implemented conditional traces to support conditional execution without host synchronization (Section 5, Section 6). In addition, we implemented many optimizations for existing kernels and implemented new kernels to execute operations more efficiently.
6. **Compatibility with existing serving and training software.** The stack supports SGLang-JAX, SGLang-TorchTPU, and TorchTitan's TPU training path. Adapters expose Tenstorrent's fused attention, loss, and mixture-of-experts kernels. Training improvements reduce the memory used by gradients, optimizer steps, and loss computation. Libtt's recurrent kernels maintain state across tokens and support speculative decoding (Section 9, Section 10).

We begin with the hardware and its upstream software foundations (Section 2, Section 3). We then describe libtt's design: the shared library, the compiler, the runtime, and execution across multiple chips (Section 4 to Section 7). Next we evaluate the stack on real workloads, covering framework integration, training, and language-model serving with performance results (Section 8, Section 9, Section 10). We close with framework compatibility results (Section 11) and a conclusion with directions for follow-up work (Section 12). The implementation is available in the project repository [12].

## 2. The Machine

A Blackhole chip has 120 small computers, which Tenstorrent calls Tensix cores, laid out in a grid [1]. Each has 1.5 megabytes of memory, five 32-bit RISC-V processors, a unit that multiplies matrices, a unit that does arithmetic on vectors, and connections to two networks that join it to its neighbors [13]. Beside the grid are 32 gigabytes of GDDR6 DRAM, which any core can read and write over the networks (Figure 2). There is no cache shared between cores. The run-time system gives 110 of the cores to programs.

A card holds one or two chips and plugs into a PCI Express slot. Most measurements in this paper were made on a TT-QuietBox 2, a workstation with four chips on two cards, in which the chips are joined by 800 Gbps Ethernet. Tenstorrent also offers datacenter servers with 32 Blackhole chips, which can be linked over the same Ethernet interconnect in various topologies. Table 1 compares Blackhole with GPUs and TPUs.

### 2.1. Tiles

Tensors are usually stored as tiles of 32 by 32 numbers. The last two dimensions are rounded up to multiples of 32, with unused entries filled by padding. For large weight matrices, this overhead is usually small. During autoregressive decoding, however, each request contributes only one token per step, so small batches often leave most of a tile's 32 rows unused. A single activation row of 4,096 numbers occupies 128 tiles, with 31 of every 32 entries devoted to padding. Conventional GPUs can operate on these short rows without requiring the same 32-row padded storage layout.

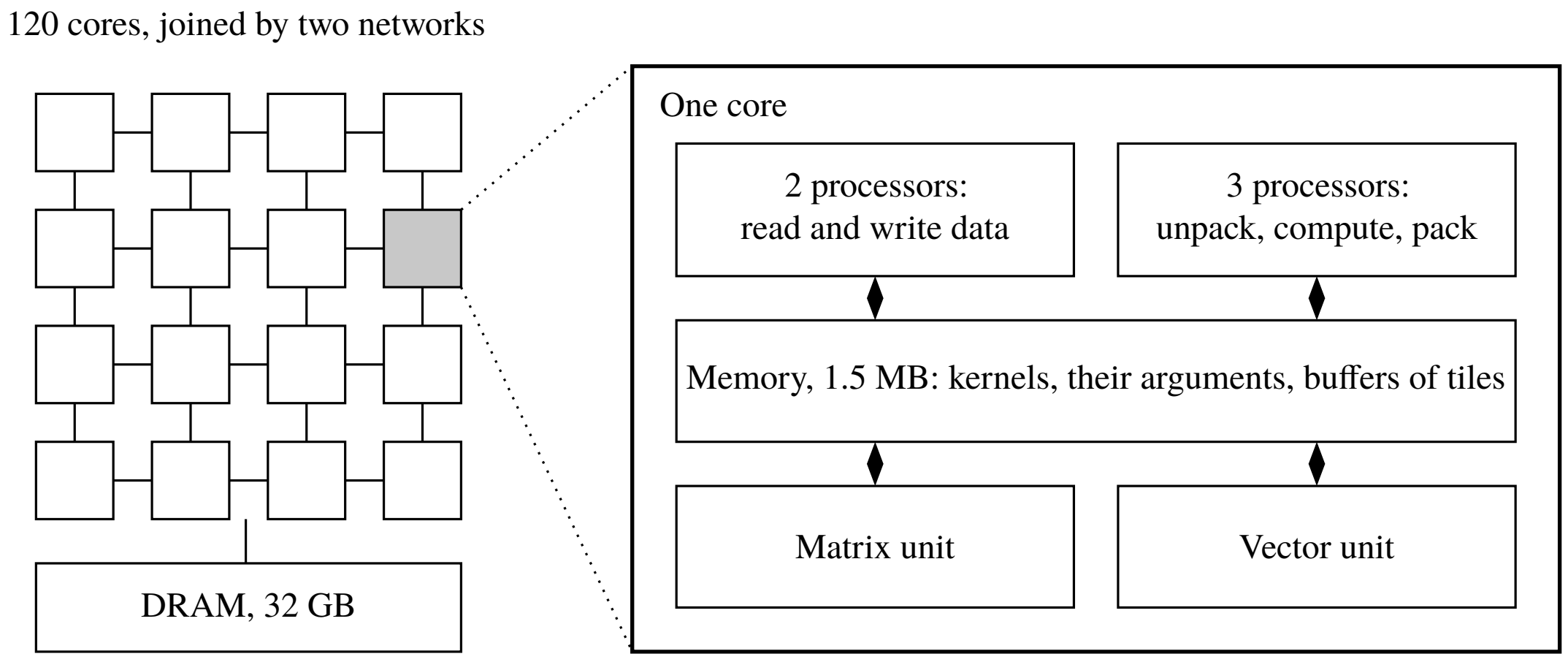

**Figure 2.** A Blackhole chip is a grid of cores beside a large memory. Each core is a small computer with five RISC-V processors, a vector and a matrix unit, and a small amount of local SRAM.

Tensors can also be stored in row-major form without this padding, but the chip's tile-based matrix units cannot operate on that layout directly.

**Table 1.** Architectural features relevant to libtt, based on vendor and compiler documentation [13], [14], [15], [16]. GPU details vary by architecture; the Tenstorrent column describes Blackhole.

| Aspect | GPU (NVIDIA, AMD) | TPU (Google) | Blackhole (Tenstorrent) |
|---|---|---|---|
| Compute units | Thread-based arithmetic and specialized matrix units | Matrix, vector, and scalar units | Matrix and vector units in each Tensix core |
| Execution | Hardware schedules groups of threads: warps or wavefronts | Compiler schedules operations and overlaps transfers | RISC-V controllers run cooperating compute and data-movement programs |
| On-chip memory | Registers, caches, and software-managed shared memory | Registers and software-managed vector/scalar memory | Registers and software-managed L1 SRAM in each core |
| Data movement | Loads and stores through caches; explicit transfers to shared memory | Scheduled transfers between HBM and on-chip memory | Explicit transfers between DRAM and cores over the on-chip network |
| Tiling | Kernel and matrix instructions determine tile shapes, tensors often row-major in main memory | Typically 8 × 128 tiles, with extra packing for narrower types [17] | Usually 32 × 32 tiles, with subtile support; tensors usually kept tiled in main memory for performance |
| Main memory | HBM or GDDR | HBM | GDDR6 |
| Software stack | CUDA or ROCm, with framework backends | XLA and libtpu, with framework backends | TT-Metalium and TTNN; libtt is a drop-in replacement for libtpu, integrates JAX and PyTorch |
| Compilation and replay | Framework graph compilation; CUDA/HIP graphs capture and replay launches [14], [18] | Framework tracing and XLA compilation; reuse compiled executables [19] | Libtt compiles framework graphs and uses device traces to replay their commands [12] |

Numbers in a tile are often 16-bit floats in the BF16 format or 8-bit floats in FP8 format with explicit scales. Numbers can also be stored in *block* formats, in which sixteen neighboring values share one exponent and keep 8 or 4 bits each (BFP8 and BFP4). The hardware also supports 32-bit float and various different integer formats.

### 2.2. The five RISC-V processors

Each core has five small processors, called baby cores by Tenstorrent. They are 32-bit, in-order, single-issue RISC-V processors (RV32IM plus a few small extensions). They perform none of a model's arithmetic; they decide what data moves and what the arithmetic units do next.

Two of them move data. Either can instruct either network to copy bytes between DRAM, this core's memory, and another core's memory. Usually one runs a *reader*, which brings tiles in, and the other a *writer*, which sends results on.

The other three drive the arithmetic. The first directs the unpacker, which loads tiles from memory into the arithmetic units' registers. The second issues instructions to the arithmetic units. The third directs the packer, which stores results back to memory.

All five share the core's 1.5 megabytes, which hold their code and the buffers between them. Each buffer is a queue of tiles: the reader fills one and the unpacker empties it, and the packer fills another and the writer empties it. The stages therefore overlap: the next tile arrives while the units work on the current one.

### 2.3. The two arithmetic units

A core has two arithmetic units: a matrix unit that is fast but coarse, and a vector unit that is general but slow. For each operation, libtt must choose between them and select the precision settings.

The matrix unit multiplies tiles. One instruction multiplies an 8 by 16 block by a 16 by 16 block, which is 2,048 multiplications and as many additions, and the unit issues one such instruction per cycle [20]. A 16 by 16 block is called a *face*, and a 32 by 32 tile consists of four faces, so multiplying two tiles takes 16 instructions. The unit also adds, subtracts, and multiplies element by element, and takes the maximum of a column.

The unit is fast because its datapath is narrow. Operands are read through 19-bit registers, which is enough for BF16 and the block formats but not for a 32-bit float, which loses the low 13 bits of its fraction on the way in [20]. The multipliers are narrower still, so a single pass uses only the leading bits of each operand. Using the remaining bits takes further passes, called *fidelity phases*, each as costly as the first. For BF16, two phases cover nearly all of the bits and four cover all of them.

The vector unit handles everything else, such as the nonlinear functions in activations, norms, and softmax. It operates on 32 values at a time, each a 32-bit float or integer, and it reads and writes the matrix unit's results directly, so a product can feed a function of that product without leaving the core. It is also much slower: one instruction processes 32 values, where a matrix-unit instruction processes 2,048.

### 2.4. Programs

Work on the chip is expressed as kernels. A *kernel* is a small C++ program. In the usual arrangement, one processor runs a *reader*, which fetches tiles from DRAM or from another core into buffers in the core's memory; three run a *compute* kernel, which unpacks tiles into the arithmetic units, operates on them, and packs the results; and one runs a *writer*, which sends the results on. The kernels on all the cores that take part, together with their arguments, make up a *program*. One operation of a model, such as a matrix product or a norm, is ordinarily one program.

Kernels are compiled on the host by a RISC-V compiler the first time they are needed, and the compiled binaries are cached on disk. The host does not start programs on the cores itself. Instead, it writes commands into a queue, and two kernels that are always running on the chip, a *prefetcher* and a *dispatcher*, read the queue, hand each core its kernels and arguments, and signal it to start.

A sequence of commands can also be recorded on the chip and replayed later with a single command. Such a recording is called a *trace*.

### 2.5. Impact on the software stack

Three characteristics of the hardware and the PyTorch and JAX workloads we run on top guide libtt's design.

*Starting a program is not free*, so libtt fuses programs. A small program, one that adds two tensors, say, takes about 5 microseconds on all 110 cores and about 2 on only the cores with work. That is small, but a language model runs a great many programs during decode: Qwen3.8-27B on four chips runs more than 1,000 for each decode step. So libtt fuses operations to do the same arithmetic in fewer programs, and launches each one only on the cores that have work. It also uses tracing to avoid host dispatch overhead when launching programs.

*A row is a bad shape for a tile*, so libtt moves only the rows that hold data. Many intermediate tensors during small-batch decoding contain only one row, or a few. Reading and writing whole tiles can therefore move up to 32 times as much data as the computation needs. Some of libtt's kernels differ from the ones they replace in touching only the rows that hold data. Sometimes we use different logical and physical tensor dimensions to reduce wasted space by padding. For example, a single large vector of size $n$ might be stored as a $\left(\frac{n}{32}, 32\right)$ shape tensor.

*Speed is limited by memory*, so libtt keeps the memory busy with weights and with little else. To produce a token, a dense model reads nearly every one of its weights, and in a decoding step the time of a matrix product goes to streaming them. Libtt's kernel for this product gives every core weights to read, reads on both of a core's networks at once, and spreads the reads over the banks of memory. With several chips, each holds a share of the weights and all read at once. Intermediate results stay on the chip: a fused program passes them from core to core instead of through memory.

These choices must also preserve the program's numerical requirements. The compiler selects arithmetic units and accumulation precision accordingly, rather than relying on the fastest defaults.

## 3. Upstream Components

Libtt builds on three upstream Tenstorrent projects: tt-xla supplies the PJRT plugin [9], tt-mlir the compiler [10], and tt-metal the runtime, driver, and base kernel library, including the TTNN operation library [11].

TTNN's documented path for running a model is to port it: rewrite its operations against TTNN's Python API and choose tensor layouts, memory placement, and core configuration by hand [21]. This yields a separate implementation of each model. Libtt instead accepts framework programs as they are and makes those choices in the compiler and runtime, without the need for modifications or separate TTNN Python bindings (Section 4).

The upstream route for frameworks, tt-xla, was not sufficient for existing workloads. For JAX, missing lowerings and incorrect operation semantics were pervasive, and for PyTorch, PyTorch/

XLA cannot run unmodified PyTorch code. The indexing, numerical, reduction, and control-flow changes in Section 5 close the JAX gap, and Section 11 reports the resulting coverage. For PyTorch, libtt uses TorchTPU and its native eager and `torch.compile` execution model (Section 8.2).

## 4. The Library

### 4.1. One library

Everything is in one shared library, `libtt.so`. It contains the PJRT plugin, the compiler with LLVM's MLIR libraries [22] beneath it, StableHLO and the Shardy partitioner [23], the run-time system and its library of operations, and the driver that talks to the chip, all linked statically via Bazel. It exports three symbols specified by PJRT, and a linker script hides the rest. The compiler's output calls TTNN's C++ operations directly. The Python code shipped with libtt registers the plugin and adapts framework and server interfaces (Section 8, Section 10).

### 4.2. The compiler

Kernels are compiled on the user's machine, so the library has to bring a compiler with it. Embedded in `libtt.so` is a compressed archive that holds a RISC-V GCC, the sources of every kernel and of the firmware, and the linker scripts. When the library is loaded it unpacks the archive into a temporary directory named after the archive's hash, and points the run-time system at it. A file lock makes this safe when several processes start at once. The compiler and runtime are compressed aggressively, so the full release build of `libtt.so` is less than 90 MB.

### 4.3. Patches

Making the upstream stack support existing framework programs required substantial changes throughout the system. As of the date of writing, libtt carries nearly 400 patches across the open-source Tenstorrent software stack with over 50,000 lines of change, together with approximately 15,000 lines of additional source code. These changes span the PJRT interface, compiler, run-time system, and device kernels: they add missing functionality, correct operation semantics, extend numerics type support, and improve performance significantly. The build pins the upstream revisions and combines them with these extensions into a reproducible stack.

Many of these changes address general limitations of the underlying software and would be natural candidates for upstream inclusion. In particular, correctness fixes, numerics type support, and general-purpose compiler and kernel improvements could benefit other users of Tenstorrent's stack. Contributing these changes upstream would broaden their availability and reduce the work of maintaining them separately. Libtt could then concentrate more of its development on framework integration and optimizations that build on those shared capabilities.

## 5. Compiling

### 5.1. The Compilation Stages

A framework passes the plugin a StableHLO program along with compile options. The plugin processes it in four stages (see Figure 3).

First, if the program is to run on multiple chips, the Shardy partitioner decides how each tensor is divided among them, and the program is rewritten to describe the work of a single chip (Section 7).

Second, StableHLO is lowered to TTIR, a tensor-level dialect that has no notion of tiles. Types the hardware does not support are legalized at this stage: complex numbers become pairs of floats, 64-bit floats become 32-bit, and signed 8- and 16-bit integers are held in 32 bits. FP16 and the supported FP8 formats keep their own tensor types, and operations without a native FP8 implementation compute on widened values (Section 5.3).

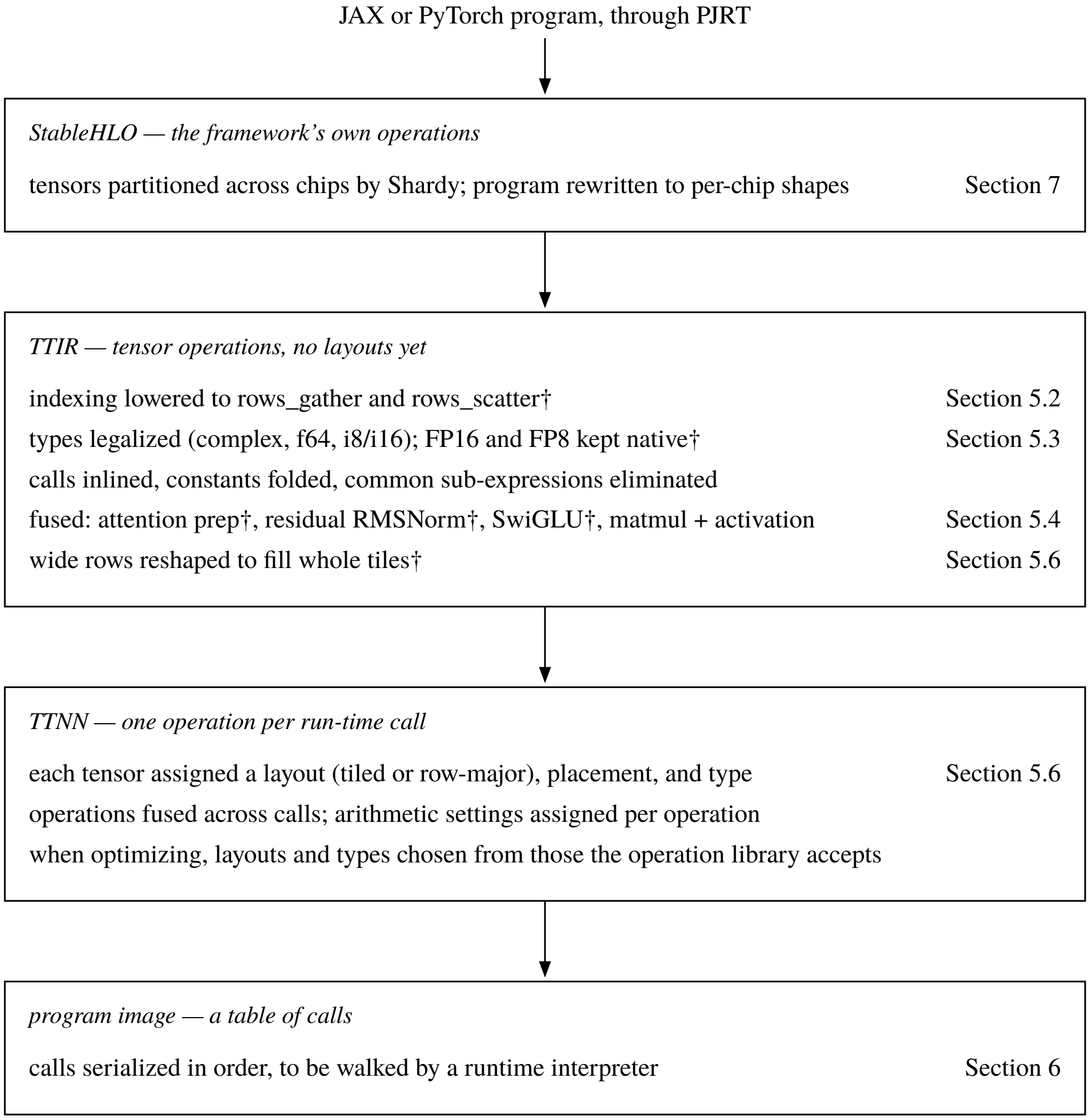

**Figure 3.** How the compiler rewrites a program, from StableHLO to the table of calls that the run-time system walks. † marks what libtt adds; the rest is extended from upstream tt-mlir.

Third, TTIR is lowered to the TTNN dialect, in which each operation is a single call into the runtime's operation library and, as a rule, a single program on the chip. Each tensor is assigned a layout (tiled or row-major), a placement (host or device), and a type, and each operation is assigned its arithmetic settings.

Finally, the result is serialized as a table of calls, which an interpreter in the runtime walks. No machine code is generated for the host.

The compiler defines three optimization levels, of which two are in use. At the lowest, every operation uses the highest-precision arithmetic the matrix unit offers. At the next, an optimizer queries the operation library for the layouts and types each call accepts and selects among them. SGLang-JAX compiles at the second level; torch_tpu at the lowest.

The upstream compiler already includes all four stages. Libtt's additions are described in the sections that follow.

### 5.2. Matching framework semantics

A framework's test suite is largely a list of small questions about what an operation means. Does a NaN compare equal to itself? What does a scatter do with an index past the end? Which of two equal keys does a sort place first? The frameworks answer each of these, and much of our work went into making libtt give the same answers.

Table 2 lists the behaviors libtt matches, grouped by area, and each entry stands for one or more changes to the compiler, the operation library, or a kernel.

**Table 2.** Framework behavior that required changes to the upstream stack

| Area | Framework behavior that libtt matches |
|---|---|
| Comparison | A NaN is not equal to itself, so floating-point self-comparisons are no longer folded. Total-order comparisons, which JAX's sort and `searchsorted` use, place −0 below +0 and NaNs at both ends. |
| NaN and signed zero | Cumulative maxima and minima propagate NaN, and the `sign` of a NaN is NaN. Negation and `log1p` keep the sign of zero. Moving or repacking a 16-bit float preserves its bits, where the default path flushed subnormals, folded −0 into +0, and rewrote NaNs. |
| Integer arithmetic | Remainder truncates. Powers wrap modulo $2^{32}$, as multiplication does, and unsigned negation is modular. High-word multiplication, arithmetic right shift, leading-zero and population counts, and integer dot products are supported. Integer select, clamp, and permute stay in integer arithmetic. |
| Floating-point types | IEEE FP16 is a type distinct from BF16 from the PJRT buffer down to the kernel. `float8_e4m3fn` and `float8_e5m2` remain FP8 in buffers, compiler types, and the runtime, and layout changes, gathers, and copies preserve their bytes. Operations without a native FP8 implementation compute on widened values, and the quantization kernels return FP8 tensors directly. |
| Narrow integer types | Signed 8- and 16-bit integers are held in 32-bit integers, and integer operations on them stay in integer arithmetic. |
| Conversion | Float-to-integer conversion truncates toward zero. Integer narrowing keeps the low bits instead of saturating. A chain of explicit conversions keeps each intermediate rounding. Unsigned values keep their value: a `uint8` 255 is not −1. Bitcasts copy raw bits, and `reduce_precision` rounds as XLA does. |
| Sorting | Equal keys keep their order in a stable sort. NaNs sort above +∞ as one tie class. Integer keys and values sort as integers or, when narrower than 32 bits, exactly through FP32. Non-strict comparators are accepted. |
| Argmax, argmin, top-k | Ties go to the smaller index. An initial value that is not the identity takes part in the reduction. Top-k returns its results in order. |
| Reductions | Integer sums, maxima, and minima are computed exactly in integer arithmetic instead of passing through FP32, which rounds above $2^{24}$. `and`, `or`, and `xor` reduce bitwise on integers and logically on booleans. Variadic reductions, which return several results at once such as a minimum and a maximum together, are supported, as are cumulative sums, maxima, and minima and general scans. |

| Area | Framework behavior that libtt matches |
|---|---|
| Convolution and pooling | Convolution keeps StableHLO's cross-correlation convention where the operation library reverses the weights. Negative padding crops the input, and reversed windows and input dilation are handled. A single-channel convolution is no longer taken for a depthwise one. Padding folds into a pooling only when it supplies the reduction's initial value. |
| Indexing | Gathers, scatters, and slices at computed positions take multi-component indices and windows. An out-of-range gather index is clamped, and an out-of-range scatter update is dropped. A scatter may add, multiply, or take the maximum or minimum, and applies duplicate updates in order. All of these lower to two new operations, `rows_gather` and `rows_scatter`, which treat the operand as a table of rows and apply the out-of-bounds rule in the kernel, replacing the runtime's embedding lookup, gather, and scatter. Only the indices and updates are rearranged, so an embedding lookup does not copy its table. |
| Control flow | A while loop repeatedly invokes an outlined program and updates its state. Conditionals and multi-way branches may nest, and may run inside a trace (Section 6.1). |
| Complex numbers | Held as pairs of floats, with multiplication, division, absolute value, equality, selection, transposition, reductions, and contractions such as dot products and convolutions. |
| Empty tensors | Concatenations and convolutions with empty operands fold to constants, where the row-major concatenation divided by a zero dimension. Zero-sized tensors pass through layout conversion and buffer casts. |
| Random numbers | StableHLO's random-bit generator runs on the device with Threefry and Philox and keeps its 64-bit state, so seeded draws repeat in eager and compiled code. |
| Other operations | Triangular solves expand into batched matrix products, as XLA expands them for TPUs. Batch normalization returns its training statistics. Deprecated operations that older frontends still emit are legalized. |
| PJRT interface | Buffers may be donated, borrowed from host memory with arbitrary strides, or hold 64-bit integers. Executables serialize for the frameworks' persistent caches. A custom call may return several results. Failures report their cause through the interface instead of escaping as exceptions. A process may create a new client before it destroys the old one. |

## 5.3. Numerics

### *5.3.1. FP16 and FP8*

Supporting a numeric type requires agreement at every layer, from the PJRT buffer down to the kernel. Libtt adds IEEE FP16 as a type distinct from BF16 throughout this path. Operations that only move values preserve their bit patterns, including signed zeros and subnormals, by copying the underlying bits rather than passing them through floating-point arithmetic.

The FP8 formats float8_e4m3fn and float8_e5m2 likewise remain FP8 in PJRT buffers, the compiler's tensor types, and the runtime. These are per-element floating-point formats, distinct from the BFP8 block format described in Section 2. Layout changes, gathers, copies, and other data movement preserve their bytes, and the quantization kernels return FP8 tensors directly.

Libtt recognizes PyTorch's scaled matrix products and uses the matrix unit's native FP8 support for supported operand pairs, with FP32 accumulation [12]. Activation quantization runs in one device program or is fused into the preceding normalization, and scales are applied within the matrix-product kernel. These changes reduce the overhead around FP8 linear layers while preserving their numerical behavior; serving results are given in Section 10.5.

#### *5.3.2. FP32 and integers*

Highest-precision FP32 matrix products run on a new vector-unit kernel, which retains FP32 operands and accumulation instead of the matrix unit's reduced operand precision, at a substantial cost in speed. The compiler preserves this request through optimization. PyTorch selects this kernel when highest precision is requested; its default uses the matrix unit's reduced operand precision. Other changes keep 32-bit integer collective sums and cumulative sums in integer arithmetic, avoiding the loss of exactness from conversion to floating point. These sums wrap on overflow as the framework requires.

### 5.4. Fusion

A model written in a framework is a graph of small operations. An RMS norm, for example, is a square, a mean, an addition, a reciprocal square root, and two multiplications. Running these separately means starting several programs and repeatedly reading and writing intermediate results. The compiler recognizes such groups and replaces them with one kernel that does the work in a single program. This is called *fusion*. The same computation can look different in JAX and PyTorch, so the compiler recognizes both forms. Libtt adapts these patterns for activation functions, normalization, rotary embeddings, and attention preparation, preserving the original casts and rounding. The fusions remain available at the low optimization level used by `torch_tpu`. Table 3 gives some examples.

**Table 3.** Selected fusions used by libtt [12]. † marks fusions added by libtt; unmarked entries adapt upstream fusions to a broad range of real-world PyTorch and JAX patterns and data types.

| Operations | Combined work |
|---|---|
| Matmul + bias or activation | Apply bias, SiLU, or GELU before storing the result. |
| Convolution + activation | Apply ReLU or SiLU inside the convolution. |
| Elementwise arithmetic + activation | Combine activations with addition or multiplication. |
| Matrix products sharing an input | Use one wider product, with weights prepared once. |
| Softmax | Subtract the maximum, exponentiate, sum, and divide. |
| Residual addition + RMS norm† | Add the residual and normalize in one program. |
| SwiGLU† | Fuse paired projections, SiLU, and the gating multiply. |
| Residual RMS norm + FP8 conversion† | Produce FP8 values and row scales during normalization. |
| Attention input preparation† | Split heads, normalize queries and keys, and apply rotary embeddings. |

A fused kernel must be both correct and fast. Fewer programs do not by themselves guarantee a speedup, because a kernel can still waste time reading padding or moving data between cores. Fusion must also preserve the rounding of the operations it replaces: moving arithmetic between the matrix and vector units, or removing an intermediate cast, can change the result. Several of libtt's fused kernels reproduce every rounding step, so their results match the unfused operations bit for bit.

Some improvements require no new kernels and come entirely from compiler rewrites. A matrix product that yields one number per batch becomes a multiplication followed by a sum. A program that keeps only the largest softmax probabilities and renormalizes them instead selects the largest inputs first and applies softmax to those alone. Both rewrites reduce the work that existing kernels perform.

### 5.5. Extending the framework with custom calls

Some operations fall outside what StableHLO's primitives express efficiently. For these, the framework provides an extension point at runtime: a JAX program invokes a *custom call* by name, and the compiler lowers it directly to a libtt kernel.

```
cache = jax.ffi.ffi_call(
    "tt.paged_update_cache", result_shape,
    input_output_aliases={0: 0},
)(cache, new_kv, positions, pages, share_cache=True)
```

Custom calls cover attention over a paged cache, recurrent layers, mixture-of-experts layers, and row quantization. For training, they add forward and backward attention and fused loss calls that run tt-train's kernels (Section 9). Keyword arguments are passed via the call's typed attribute dictionary. The alias allows the result to overwrite the first argument, which is how a cache is updated in place.

### 5.6. Tiles and precision

The tiled storage described in Section 2 affects far more than matrix multiplication. A reshape, slice, or gather can force data to move between tiles, and a tensor with only a few rows is mostly padding. These costs recur throughout a model, so layout selection is a major concern for the compiler.

Libtt keeps data in layouts that neighboring operations can consume directly, which avoids unnecessary layout conversions, and it moves elementwise work to shapes that occupy fewer tiles. For wide rows, it regroups data into fuller tiles for reductions or uses kernels that skip the unused parts of tiles, reducing both memory traffic and work spent on padding [12]. The rotary-embedding example in Section 9 shows how a dimension of two occupies 32 columns in tiled storage, and how a rewrite avoids that padding.

These layout choices must also respect the numerical requirements in Section 5.3. Libtt uses vector-unit arithmetic and 32-bit accumulation where needed, and its optimizations preserve these choices instead of falling back on the runtime's defaults.

## 6. Executing

### 6.1. Traces

A decode step can contain hundreds or thousands of small device programs, and dispatching each one from the host leaves the chip idle while it waits for work. Upstream TT-Metalium offers traces, recorded sequences of device commands that replay with a single request [11]. Libtt greatly extends this mechanism so it supports the dynamic execution patterns real PyTorch and JAX workloads exhibit. Short eager operations run without trace capture, while larger programs can use replay.

*Memory.* Recorded commands refer to fixed addresses, so libtt tracks the memory each trace uses and checks that replay cannot overwrite live tensors. Libtt also supports allocations inside a trace and reserves the DRAM used by temporary tensors, so unrelated allocations rarely force a trace to be recorded again. Under memory pressure the reservations are released, and the runtime re-records any trace that needs it.

*Inputs.* Tensors already on the device stay there during replay. They are read in place when possible, or copied into fixed device buffers when their addresses change. Even inputs the compiled program expects from the host can be staged this way, so a result never has to return to the host just to feed the next compiled function.

*Branches.* Libtt adds conditional commands to the command processor, and the compiler records supported branches inside the trace. The device reads the condition from memory and skips branches that are not taken, so nested conditionals such as the token sampler run with no round trip to the host.

Every branch's kernels must be compiled and loaded before capture, so libtt adds a preparation mode that does this without executing them. Preparing an unselected branch therefore leaves the model's state unchanged.

### 6.2. Host overhead

Traces cut the cost of dispatching device programs, but the host still prepares arguments, transfers inputs, and receives results. Libtt shrinks this work and overlaps it with device execution wherever dependencies allow [12].

*Caching.* Most of what is needed to launch a compiled program is the same from call to call. Libtt caches input layouts and the results of input analysis, builds sharding information only when a new runtime tensor requires it, and starts trace replay directly without constructing another program executor. Compiled programs can also be saved and loaded for `torch.compile`'s persistent cache, provided the same libtt build is used.

*Data movement.* Small transfers such as token IDs and positions are issued from the calling thread, since handing them to a thread pool can cost more than the copy itself. Larger host copies use contiguous blocks and, for supported transposes, vector instructions, and read buffers are no longer zeroed before being overwritten. Where supported, tiled results are converted to row-major on the device before transfer, which reduces both the bytes read and the CPU work.

*Overlap.* Libtt queues result transfers in program order and releases submission locks before waiting for completion, so a serving system can read finished tokens while it prepares and submits later work, with no gaps between device computations.

### 6.3. Memory

Weights are large, and moving them accounts for most of a server's startup time. Libtt supports memory-mapped checkpoints, without first copying each tensor, and can convert weights to their block format on the chip instead of on the host.

## 7. Distributed Execution Across Chips

Libtt runs a single program across multiple chips. It organizes the chips into a logical *mesh*, places each array by dividing it across one or more mesh axes, replicating it across others, or holding it on one chip, and provides collective operations that exchange or combine data when a computation needs values held on other chips.

This supports many different ways of dividing work, including the following ones: *Data parallelism* assigns different examples to different chips, usually with a copy of the model on each. *Tensor parallelism* divides the tensors and computations within a layer [24]. *Expert parallelism* places the experts of a mixture-of-experts model on different chips. *Fully sharded data parallelism* distributes parameters, gradients, and optimizer state across chips and gathers parameters when needed. *Pipeline parallelism* assigns successive groups of layers to different chips and passes intermediate results between them. *Sequence parallelism* divides some intermediate activations along the sequence dimension, and *context parallelism* also distributes attention over parts of a long sequence. TorchTitan's four-chip training runs are a concrete example of fully sharded data parallelism (Section 9).

Libtt implements the local computation and communication that the framework's partitioning requires [12]. The chips can be driven from a single controller or from one process per chip, as described below.

### 7.1. Single controller

A JAX program states how its arrays are divided by annotating them, and the Shardy partitioner propagates the annotations through the program. With `shard_map`, the programmer can instead write the per-chip computation explicitly and mix it with automatically partitioned code. Libtt supports both forms and their combination. After partitioning, the compiler rewrites every shape to the piece one chip holds, and whatever must cross between chips appears as an explicit *collective*: an all-reduce that sums, or takes the maximum or minimum of, the values the chips hold; an all-gather that concatenates them; or a send from each chip to another.

A single controller drives all the chips. Data crosses between them over their Ethernet links, routed by small kernels on each chip, and never passes through the host. No launcher or message-passing library is needed, so the server starts as it does on one chip, with one additional flag.

All-reduce matters most for tensor-parallel decoding, where it combines partial results within every layer. Libtt reduced its cost from two programs to one: each chip sends its rows directly into spare memory on the others, and each adds what it receives. Only rows that hold data are sent, and sums that must be exact, of integers or 32-bit floats, are computed on the vector unit.

Before libtt, a row-divided embedding table forced a choice between gathering the whole table onto every chip for each lookup and keeping a full copy on every chip. Libtt needs neither. Each chip looks the token up in its own rows and returns zeros if the row is elsewhere, and an all-reduce sums the results. Each chip therefore stores only its share of the table, so memory per chip falls as chips are added, and only one embedding vector per token crosses between chips, never the table.

In mixture-of-experts layers, each chip holds a subset of the experts and computes their contributions to the tokens routed to them, and a collective sums the partial outputs into the layer's result [12].

### 7.2. One process per chip

TorchTPU runs one process per chip, as is usual on a machine full of TPUs, and each process compiles the same program with the collectives written out. For TorchTPU, libtt follows this model. Each process opens only its own chip. The processes agree on a mesh of $N$ chips in a line and find one another through a small key-value store that the first process serves over TCP, replacing the message-passing library upstream would use. Only coordination goes through the store; tensors still cross over the chips' own links.

Collectives must write to the addresses where other chips actually hold their buffers. Upstream implementations often assume these addresses match across chips, which independent processes cannot guarantee. Libtt extends the all-gather and reduce-scatter kernels used between processes to exchange buffer addresses during their existing startup handshake, so each chip writes to the address its neighbor supplies. The small direct all-reduce instead uses buffers that its program places at matching L1 addresses.

## 8. Framework Integration

JAX and PyTorch reach libtt through PJRT. Both use the same compiler and runtime; their frontends translate framework operations into StableHLO, and libtt supplies the device implementation described in the preceding sections.

### 8.1. JAX

The Python package registers libtt as JAX's `tt` platform through a plugin entry point. Setting the environment variable `JAX_PLATFORMS=tt` selects it, and supported programs use the ordinary JAX interface, including `jax.jit`. Options can be supplied for each compilation:

```
fast = jax.jit(f, compiler_options={
    "optimization_level": "O1", "enable_trace": "true"})
```

### 8.2. PyTorch

Upstream tt-xla integrates PyTorch through PyTorch/XLA [9]. Its traditional lazy-tensor model records operations until a graph boundary and can require applications to manage execution boundaries. That is the wrong abstraction for libtt, whose goal is to run existing PyTorch software: ordinary eager execution and the standard `torch.compile` interface must remain the programming model.

Libtt therefore supports PyTorch workloads through TorchTPU, Google's PyTorch backend for TPUs, which supports both eager execution and `torch.compile` [25]. Google identifies TorchTPU as the replacement for PyTorch/XLA and recommends migrating existing workloads to it [26]. Setting `LIBTT_TPU_COMPAT=1` enables libtt's TPU compatibility layer, which lets TorchTPU discover the Tenstorrent chips and use libtt through PJRT. Multi-chip execution follows the process model in Section 7.

Software written for TPUs may also call hardware-specific kernels directly. The compatibility layer substitutes implementations when the relevant modules are imported, using Tenstorrent kernels or ordinary framework operations. This lets the supported TorchTitan and SGLang-TorchTPU paths run without editing the applications' source. The compiler and runtime changes in Section 5 and Section 6 support their mix of eager and compiled execution; Section 9 and Section 10 describe the applications.

## 9. Training

Libtt runs TorchTitan's TPU training path on four Blackhole chips without changes to TorchTitan's source [27], [28]. This requires connecting the trainer to Tenstorrent's kernels, extending the compiler for gradient computations, and reducing training's memory requirements. The results below cover full-size model throughput and a completed fine-tuning run.

The trainer uses the TorchTPU compatibility layer described in Section 8.2, with one process per chip. Fully sharded data parallelism [29] divides the parameters, gradients, and optimizer state among the chips. In these four-chip runs, each chip keeps a quarter of each, gathers a layer's weights before using them, and receives its share of the summed gradients afterward. Each chip trains on its own sequences.

### 9.1. Training kernels

The TPU trainer uses hardware-specific Pallas kernels for attention and the combined final projection and loss. Libtt connects these operations to tt-metal's existing fused forward and backward kernels [11]. Its compatibility layer adapts the trainer's calls, and the compiler carries them through to the kernels. For Qwen3.5's gated RMS norm, libtt selects TorchTitan's ordinary PyTorch implementation.

The fused loss avoids storing a full tensor of vocabulary-wide logits. It processes tokens in chunks and recomputes each chunk's logits during the backward pass. Together with fused attention, this allowed Qwen3-8B to train with four sequences per chip instead of one, increasing throughput from 287 to 435 tokens per second per chip.

Libtt also extends the attention kernels for packed documents. A token may attend only to earlier tokens in its own document, leaving much of the attention matrix unused. The kernels recognize this structure and skip the masked blocks. For Llama 3 8B, this reduced a training step from 30.5 to 22.0 seconds.

### 9.2. Backward pass

Training exposes costs that inference never sees, and libtt targets four of them:

*Balancing embedding updates.* The gradient of a token embedding adds each token's gradient into the corresponding row of the embedding table, and real text uses some tokens far more often than others. Libtt distributes the rows among cores in round-robin order, which balances these updates: Qwen3-1.7B's embedding gradient takes about 30 milliseconds instead of 238.

*Removing differentiation overhead.* Libtt removes some of the work that automatic differentiation generates. The compiler combines gradients of adjacent slices into a concatenation, avoiding large zero-filled intermediate tensors, and expands the triangular solves in Qwen3.5's recurrent layers into matrix products. Optimizer changes remove unnecessary broadcast constants, and matrix-product kernels use blocking suited to the large shapes found in training.

*Eliminating tile padding.* Tiled storage matters throughout the backward pass. Llama 3's rotary-embedding expression creates a trailing dimension of two, which occupies 32 columns in tiled storage, a sixteenfold padding cost. Libtt rewrites the operation to work on separate halves of each head, which avoids that cost and preserves the values and gradients. This is one example of the layout work described in Section 5.

*Freeing collective memory.* Collectives need temporary storage, and libtt reduces it: Large reduce-scatter operations run in pieces so they no longer need one large contiguous allocation. This makes more of each chip's memory available to the model, so it can use larger batches.

### 9.3. Throughput and CPU checks

Table 4 reports throughput for full-size models on four chips.

The checks against a CPU are separate from these full-size throughput measurements. Tested small models and selected layers are trained for a few steps from the same saved starting point on both the CPU and the chips. They include TorchTitan's debug models and the first two layers of Llama 3 8B with its full vocabulary and 8,192-token sequences. In these checks, the first losses agree to three digits or more, and subsequent losses differ by at most about one percent. Different kernels, rounding, and summation order can produce small differences that accumulate across updates.

**Table 4.** Training on four chips. Parameters, gradients and AdamW's state in BF16. Text from C4 (for the Llama models, TorchTitan's small sample of it), packed into the sequences. For the Llama models, the mean of the steps after the first, which compiles the programs.

| Model | Tokens per chip | Seconds per step | Tokens per second per chip |
|---|---|---|---|
| Qwen3-8B | 4 × 2,048 | 18.8 | 435 |
| Qwen3.5-9B | 4 × 2,048 | 30.1 | 272 |
| Llama 3 8B | 8,192 | 22.0 | 372 |
| Llama 3.2 3B | 8,192 | 13.8 | 595 |
| Llama 3.2 1B | 8,192 | 5.82 | 1,407 |

The DeepSeek-V3, Qwen3 MoE, and Qwen3.5 MoE debug models run unchanged in PyTorch eager mode with 32-bit floats.

### 9.4. Supervised Fine-tuning

Qwen3.5-9B-Base is fine-tuned to call functions on ToolACE [30]: 11,300 examples whose system prompt lists functions in JSON and whose reply is the calls, as a Python list such as `[get_weather(city="Paris")]`, or a statement that none applies. The recipe adapts TorchTitan's Llama 3 fine-tuning example: a loss on the reply's tokens, whole samples of up to 2,048 tokens packed into 8,192 tokens per chip, 300 steps (1.25 passes over the data), AdamW at up to $2 \times 10^{-5}$ with 25 warm-up steps and cosine decay, and BF16 parameters, gradients and optimizer state. A fixed 2 percent is held out; weights load and save as Hugging Face checkpoints.

A step took 29.5 seconds (278 tokens per second per chip); the held-out loss falls from 0.448 after the first step to 0.302 by step 25 and 0.252 after 300 (Figure 4).

Both models answer BFCL's five single-turn categories [31] (1,240 tests), scored by its own prompt, parser and syntax-tree checker and served greedily by SGLang-TorchTPU with up to 1,024 tokens.

Fine-tuning raises the mean from 64.8 to 88.5 percent with the same prompt, and to 91.0 percent with the non-thinking prompt that matches the training format (Table 5); read leniently, the Base model reaches at most 75.8 percent.

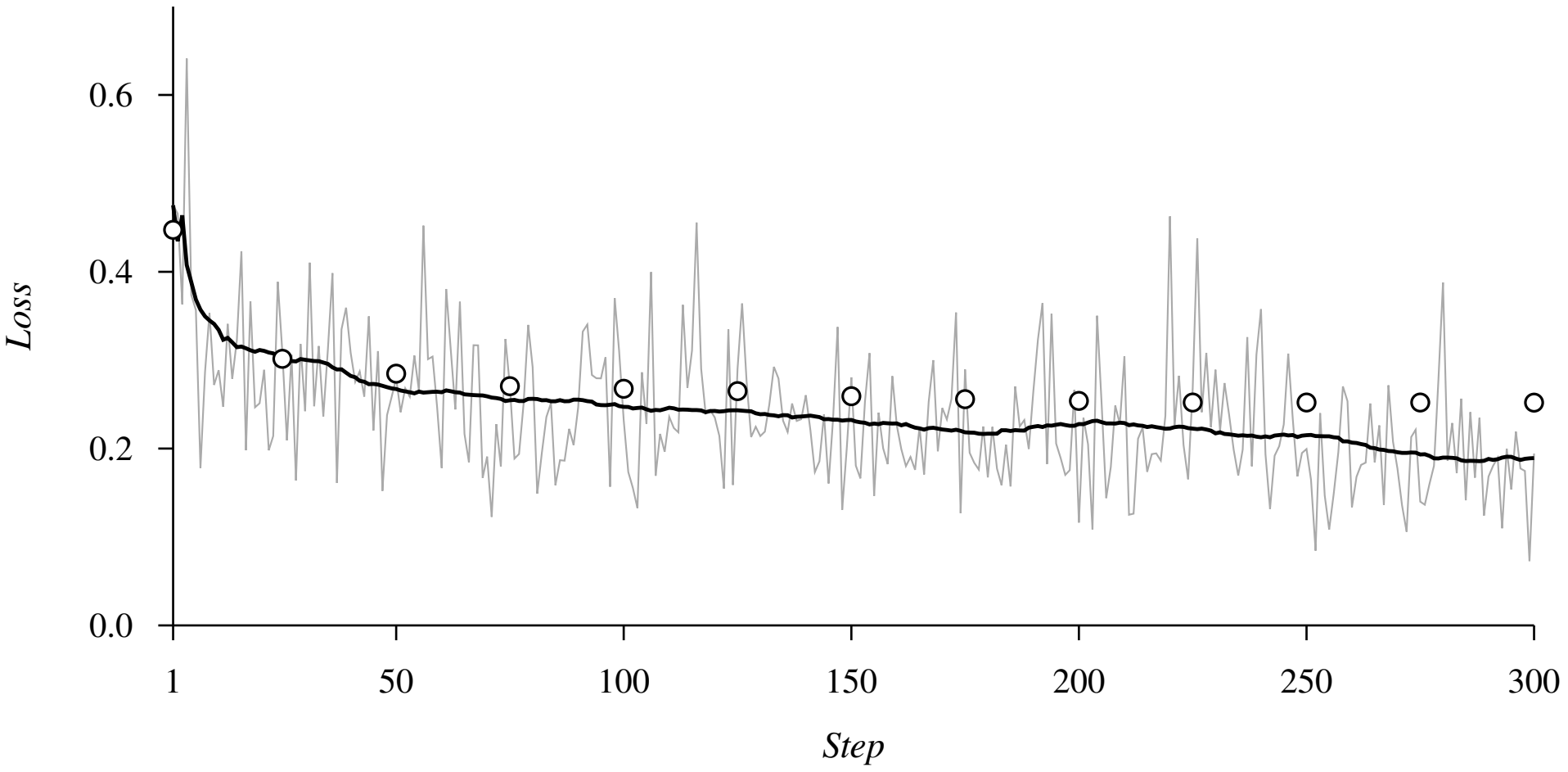

**Figure 4.** Fine-tuning Qwen3.5-9B-Base on ToolACE: training loss per step (gray), its 75-step centered mean (black) and held-out loss (circles).

**Table 5.** Function calling on BFCL (percent of tests passed) before and after fine-tuning, with the chat template's thinking and non-thinking prompts. Lenient reads any list of calls after removing copied template placeholders, a control for answer format. Mean: unweighted, over the five categories.

| Category (tests) | Base | Base, lenient | Fine-tuned | Fine-tuned, non-thinking |
|---|---|---|---|---|
| Simple (400) | 59.0 | 67.8 | 84.8 | 93.5 |
| Multiple (200) | 63.5 | 72.0 | 86.0 | 95.0 |
| Parallel (200) | 58.0 | 76.0 | 90.0 | 91.5 |
| Parallel multiple (200) | 49.5 | 70.5 | 86.5 | 83.0 |
| Irrelevance (240) | 94.2 | 92.9 | 95.0 | 92.1 |
| Mean | 64.8 | 75.8 | 88.5 | 91.0 |

We also checked the training against a CPU on MetaMathQA [32], TorchTitan's own fine-tuning example. A scaled-down run with the same starting point and data gives the same losses on the chips as on the CPU, to within 0.2 percent per step on average, in both BF16 and FP32. In serving, the chips choose the same next token as a CPU reference at about 99 percent of positions, with most differences at the limit of BF16 precision.

### 9.5. Reinforcement learning with Miles

Miles [33] is a reinforcement-learning framework that runs training and serving in one loop, which makes it a demanding test of framework compatibility. It runs its existing TPU workflow on libtt, reusing its TorchTitan trainer, SGLang-TorchTPU server and PyTorch model definitions, with no Tenstorrent-specific trainer or model port.[1]

On a TT-QuietBox 2, Qwen3-0.6B trains on two Blackhole chips with FSDP2 and generates on two with tensor parallelism. Each of 90 group-relative policy optimization (GRPO) steps samples eight answers to each of 16 GSM8K training questions [34], scores their boxed answers, updates the model, and transfers the weights. Training uses AdamW at $2 \times 10^{-6}$, BF16 computation and FP32 parameters and optimizer state; generation uses BF16 weights.

Three runs with different seeds raise accuracy on all 1,319 GSM8K test questions, with greedy decoding and a 1,024-token limit, from 58 percent at step 0 to 73.3 percent on average after 90 steps (72.4 to 74.6 percent; Figure 5). The gain is not answer format: the starting policy already boxes 99 percent of its answers. For two rollout batches, an FP32 CPU computation of the answers' log-probabilities agrees with the trainer's batch mean to within 0.0003.

Over the runs' final 15 steps, a step's generation averages 48 seconds, log-probabilities 21, training 50 and weight transfer 1.6. These runs demonstrate the complete RL loop; distributed training checkpoint and resume remain follow-up work.

## 10. Serving

Libtt currently supports SGLang-TorchTPU and SGLang-JAX for large language model serving. The server handles request scheduling and batching, while libtt compiles each step and supplies the device kernels. This section uses SGLang-TorchTPU as its main example and closes with performance measurements (Section 10.5).

The two phases of LLM serving, prefill and decode, stress different resources. Prefill processes the whole prompt at once, so it has many tokens per program and is compute-bound. Decode generates one token per request per step, so with small batches it has few tokens per program and is dominated by weight reads and program launches. The balance between the phases depends on prompt length, output length, and batch size. The server compiles programs for a fixed set of shapes, and libtt traces and replays repeated execution to cut launch overhead (Section 6).

[1] A two-file patch to Miles enables its disk-based weight transfer on the TPU platform and replaces two CUDA-specific calls.

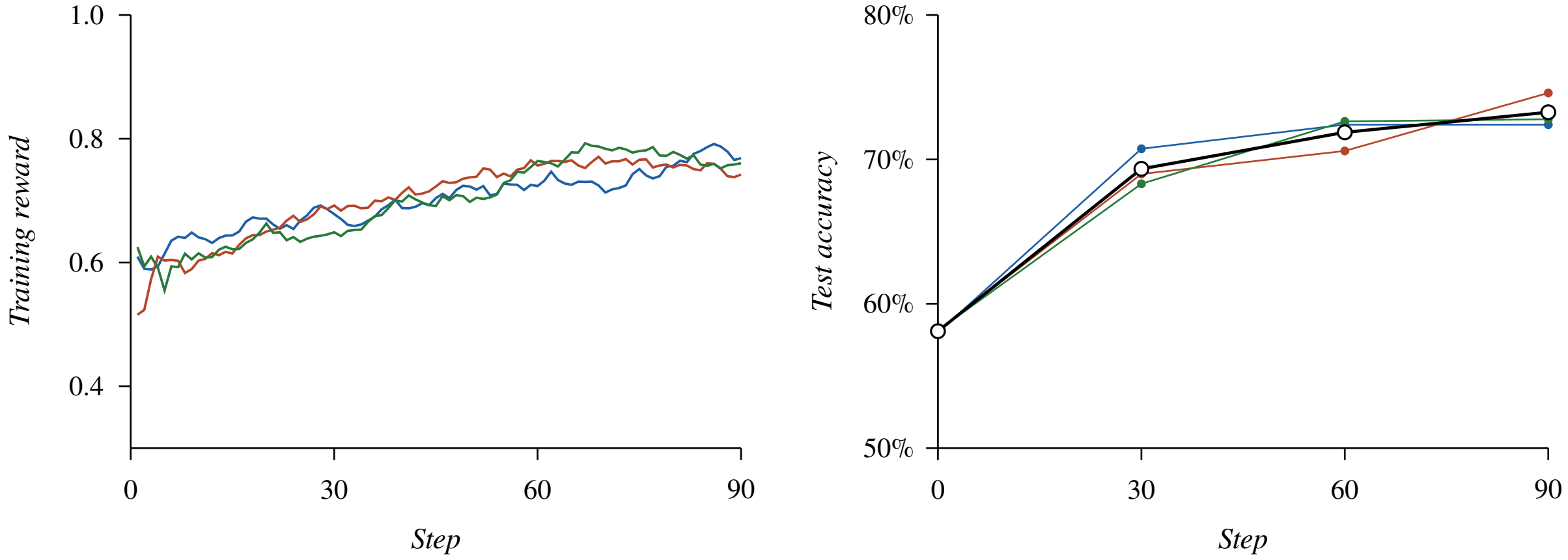

**Figure 5.** Qwen3-0.6B on GSM8K on four Blackhole chips, three runs with different seeds (color). Left: training reward per step, as trailing 15-step means. Right: test accuracy of each run and their mean (circles), with greedy decoding and a 1,024-token limit.

## 10.1. Adapting the server

SGLang-TorchTPU relies on TPU-specific kernels written in Pallas, which libtt cannot run. Libtt rewrites these kernels for Tenstorrent hardware, building on upstream Tenstorrent kernels, its own kernels, and ordinary framework operations. The compatibility layer in Section 8.2 installs the rewrites behind the interfaces the server already calls, so the server's source is unchanged on the supported paths. Table 6 summarizes the main rewrites.

## 10.2. Projections and experts

Decode-time projections are limited by how fast weights stream from DRAM, so libtt targets three places where that traffic and the surrounding work can be cut.

*Projections.* A projection multiplies activations by a weight matrix held in DRAM, and at small decoding batches its speed depends on how quickly those weights reach the cores. Libtt adapts the matrix-product kernels to keep each input row resident on the cores while the weights stream past, and to spread reads across memory banks. The bank spreading matters most for transposed PyTorch weights, whose natural read order otherwise concentrates traffic on a few banks.

**Table 6.** Serving adaptations for SGLang-TorchTPU. † Added by libtt; ‡ extended or adapted from upstream Tenstorrent.

| Operation | Device implementation |
|---|---|
| Attention and cache | Paged attention and shared updates‡; cache pool adapter† |
| FP8 projections | Quantization and scaled decode kernels†; larger matrix products‡ |
| Mixture of experts | Fused BF16/FP8 expert computation‡ |
| Recurrent layers | Decode, convolution and state-pool kernels†; chunked prefill‡ |
| Draft cache updates | Rotary/cache-write adapter† using paged updates‡ |
| Collectives | Native all-gather‡ |

*Fusion.* A feed-forward block computes two projections of the same input, applies SiLU to one, and multiplies the results. Libtt fuses the two projections and performs the remaining operations before the output leaves the core (Table 3), so the intermediates never return to DRAM.

*Mixture of experts.* Each token selects a few feed-forward blocks, for example 8 of 128 experts in Qwen3-30B-A3B, so reading only the selected experts' weights cuts memory traffic. Libtt uses Tenstorrent's fused expert kernel, including 4-bit block-format weights, and integrates it across chips: each chip computes the selected experts it holds, and their outputs are summed (Section 7). For SGLang-TorchTPU, we added support for BF16 and FP8 experts. FP8 weights are read directly and scaled inside the kernel, with one scale per expert and output column, while external inputs and outputs stay BF16. Expert selection is also kept cheap: the router finds the largest scores with a parallel scan on the data-movement processors, avoiding a full sort.

### 10.3. Attention and recurrent state

*Paged attention.* Keys and values of earlier tokens stay in a paged cache on the device. Libtt exposes the upstream paged-attention kernels for decode and chunked prefill through custom calls. It adds in-place writes of new keys and values, so the whole cache is never converted between layouts, and it divides attention work across cores.

*Cache layout.* Libtt places a head's key and value side by side in one row. Stored as two rows, that dimension would pad to 32, wasting memory and adding data movement. The server keeps its own request and page management on top of this device layout.

*Recurrent layers.* Hybrid models such as Qwen3.5 and Qwen3.8 combine attention with gated DeltaNet layers [35], which carry a fixed-size state matrix per head from token to token. Libtt supplies kernels for the decode step and the short convolution that precedes it, along with a pool of states for concurrent requests. The states stay on the device between steps.

### 10.4. Choosing and verifying tokens

*Greedy* decoding takes the highest-scoring token. *Sampling* scales the scores by a temperature, optionally filters out unlikely tokens, and draws from the resulting distribution. Libtt runs both on the device, including random-number generation. They operate on vocabulary-wide rows and can account for a substantial part of a short decode step (Table 11).

*Speculative decoding* has a cheap draft propose several tokens, which the main model then verifies together [36]. Libtt supports both a model's own draft layer and a separate draft model. Verifying candidates together can reuse work, but the speedup depends on verification cost and on how many candidates are accepted (Table 12).

Recurrent layers complicate verification, because each candidate depends on the state produced by the one before it. Libtt's kernels process several steps in one call with the state held on the device, and save the state after each candidate. Verification can then keep the accepted prefix's state and discard the rejected suffix. Attention applies the corresponding cache updates for the candidate tokens.

### 10.5. Performance

The published Blackhole serving results were measured on a TT-QuietBox 2: Four Blackhole chips on two cards, an AMD Ryzen 7 9700X, and 256 gigabytes of host memory [1]. They come from the project's pages [37], where the scripts that produce them are also to be found. All of them were taken with version 0.2.2 of libtt. Single-request figures are medians of five runs after two to warm up. With several chips the model is divided by tensor parallelism over all of them.

**Table 7.** One request, 128 generated tokens, greedy. Decoding speed in tokens per second, and time to the first token of a 215-token prompt in milliseconds. A dash means that the model does not fit on that many chips.

| | | Tokens per second | | | First token, ms | | |
|---|---|---|---|---|---|---|---|
| Model | Kind | 1 chip | 2 chips | 4 chips | 1 chip | 2 chips | 4 chips |
| *SGLang-JAX, BFP8 weights (BFP4 experts)* | | | | | | | |
| Qwen3-8B | dense | 47.0 | 81.0 | 124.4 | 113 | 76 | 54 |
| Qwen3-14B | dense | 27.0 | 49.4 | 83.3 | 179 | 110 | 76 |
| Qwen3-32B | dense | — | 22.7 | 40.6 | — | 229 | 147 |
| Qwen3.5-9B | hybrid | 43.5 | 74.7 | 117.0 | 159 | 108 | 78 |
| Qwen3.8-27B | hybrid | — | 26.7 | 46.1 | — | 271 | 179 |
| Qwen3-30B-A3B | MoE | — | 104.3 | 127.4 | — | 97 | 69 |
| Qwen3.5-35B-A3B | MoE | — | 88.9 | 109.2 | — | 141 | 104 |
| Qwen3.8-Flash-Next | MoE | — | — | 42.7 | — | — | 810 |
| *SGLang-TorchTPU, FP8 weights* | | | | | | | |
| Qwen3-8B-FP8 | dense | 44.4 | 75.0 | 116.0 | 251 | 166 | 122 |
| Qwen3-14B-FP8 | dense | 25.9 | 46.8 | 76.2 | 393 | 235 | 167 |
| Qwen3-32B-FP8 | dense | — | 22.6 | 38.8 | — | 477 | 317 |
| Qwen3.5-9B | hybrid | 37.2 | 59.9 | 86.5 | 451 | 279 | 191 |
| Qwen3.8-27B-FP8 | hybrid | — | 25.3 | 42.0 | — | 693 | 457 |
| Qwen3-30B-A3B-FP8 | MoE | — | 87.0 | 97.4 | — | 196 | 143 |
| Qwen3.5-35B-A3B-FP8 | MoE | — | 73.2 | 83.0 | — | 340 | 239 |

### *10.5.1. One request*

Table 7 gives the speed of decoding and the time to the first token. Under SGLang-JAX the matrices of weights are in BFP8, the experts of mixture-of-experts models in BFP4, and everything else in BF16. Under SGLang-TorchTPU the weights are 8-bit floats.

Speed does not necessarily grow in proportion to chips: Qwen3-8B on two chips is 1.72 times as fast as on one, and on four chips 2.65 times. The reason is that only the reading of weights is divided by four. The rest of a step is not: the small programs between the projections, the collectives, the choice of a token. Of the two dense models that fit on one chip, the larger gains more; Qwen3-14B is 3.09 times as fast on four chips.

### *10.5.2. Many requests*

Table 8 gives the total speed when several are decoded together. Since each decode step reads the weights only once, the cost of a step grows much more slowly than the number of rows in it.

### *10.5.3. Comparison with a TPU*

In Table 9 we compare the exact same SGLang-TorchTPU code in two settings: Running with the libtt runtime on a Tenstorrent Blackhole P150 card and running with Google's libtpu runtime on a TPU v6e. Per unit of memory bandwidth, one Blackhole matches or exceeds one TPU at single-request decode. The TPU is 2.8 to 3.2× faster, close to its 3.2× bandwidth advantage (1,638 GB/s versus 512 GB/s).

**Table 8.** Several requests at once on four chips: total tokens per second over all requests, prefill included. Each request has a 19-token prompt and generates 128 tokens, greedy.

| | Concurrent requests | | | |
|---|---|---|---|---|
| Model | 1 | 4 | 8 | 16 |
| *SGLang-JAX* | | | | |
| Qwen3-8B | 120 | 439 | 770 | 1,139 |
| Qwen3-14B | 81 | 302 | 541 | 902 |
| Qwen3-32B | 39 | 148 | 274 | 444 |
| Qwen3.5-9B | 113 | 392 | 660 | 999 |
| Qwen3.8-27B | 45 | 160 | 284 | 448 |
| Qwen3-30B-A3B | 121 | 424 | 707 | 1,014 |
| Qwen3.5-35B-A3B | 104 | 342 | 566 | 820 |
| Qwen3.8-Flash-Next | 39 | 100 | 145 | 170 |
| *SGLang-TorchTPU* | | | | |
| Qwen3-8B-FP8 | 106 | 367 | 580 | 750 |
| Qwen3-14B-FP8 | 71 | 249 | 422 | 609 |
| Qwen3-32B-FP8 | 36 | 129 | 219 | 344 |
| Qwen3.5-9B | 77 | 242 | 360 | 466 |
| Qwen3.8-27B-FP8 | 37 | 115 | 172 | 235 |
| Qwen3-30B-A3B-FP8 | 88 | 315 | 500 | 666 |
| Qwen3.5-35B-A3B-FP8 | 72 | 238 | 339 | 432 |

**Table 9.** Comparing LLM inference for FP8 models: 128 greedy output tokens; medians of five single requests or three batches after warm-up. Decode uses a 5-token prompt; first-token latency uses 215 tokens.

| | Decode, tokens/s | | First token, ms | | 16 requests, tokens/s | |
|---|---|---|---|---|---|---|
| Model | BH P150 | TPU v6e | BH P150 | TPU v6e | BH P150 | TPU v6e |
| Qwen3-8B-FP8 | 44.2 | 130.6 | 253.5 | 108.2 | 407.6 | 1,270.1 |
| Qwen3.5-9B | 37.1 | 119.7 | 471.0 | 123.3 | 206.3 | 1,177.6 |
| Qwen3-14B-FP8 | 25.9 | 73.5 | 395.4 | 198.0 | 251.6 | 806.1 |

### *10.5.4. Upstream Tenstorrent comparison*

We also compare with the upstream Tenstorrent software stack: a vLLM plugin that requires separate TTNN model definitions together with custom scheduling, execution, KV-cache handling and runtime compatibility patches. That stack is compared against two libtt-based stacks on the same chip: SGLang-TorchTPU and SGLang-JAX.

These comparisons are not apples to apples: the three stacks do not use the same numerics. The Tenstorrent vLLM stores most weights in the BFP8 block format (Section 2) and the rest in BFP4. SGLang-JAX stores its weights in BFP8, so it is the most comparable to the upstream stack, while SGLang-TorchTPU uses FP8 and has additional overhead from the FP8 scales. The results are in Table 10.

**Table 10.** Same prompts, 128-token outputs, and timing method as Table 9; one chip, no prefix caching or speculation. Decode uses the short prompt and first-token latency the long one. A dash marks a result that is not available. "16 requests" is total throughput at 16 concurrent requests (median of three measured batches).

| | Decode, tokens/s | | | First token, ms | | | 16 requests, tokens/s | | |
|---|---|---|---|---|---|---|---|---|---|
| Model | SGLang-JAX | SGLang-TorchTPU | Upstream | SGLang-JAX | SGLang-TorchTPU | Upstream | SGLang-JAX | SGLang-TorchTPU | Upstream |
| Qwen3-8B | 46.1 | 44.2 | 38.7 | 122.6 | 253.5 | 130.7 | 514.1 | 407.6 | 522.8 |
| Qwen3.5-9B | 43.4 | 37.1 | 22.4 | 161.0 | 471.0 | 173.5 | 410.3 | 206.3 | — |
| Qwen3-14B | 26.5 | 25.9 | 20.9 | 183.2 | 395.4 | 209.8 | 318.5 | 251.6 | 287.0 |

SGLang-JAX on libtt is as fast as or faster than the upstream software, owing to the additional fusions libtt performs in the compiler. SGLang-TorchTPU is faster than upstream at decode but slower at prefill and with larger batches, because of the additional overhead from dynamic quantization.

*10.5.5. Sampling and speculative decoding*

**Table 11.** The cost of sampling: Qwen3-8B, one request, tokens per second.

| | SGLang-JAX | | SGLang-TorchTPU | |
|---|---|---|---|---|
| Sampling | 1 chip | 4 chips | 1 chip | 4 chips |
| greedy | 47.0 | 124.4 | 44.4 | 116.0 |
| temperature 0.7 | 45.2 | 110.9 | 42.7 | 105.3 |
| and `top_p` 0.95 | 39.9 | 83.6 | 39.0 | 85.5 |
| and `top_k` 20 | 39.9 | 83.6 | 43.4 | 109.4 |
| and both | 35.6 | 66.8 | 42.4 | 103.5 |

Table 11 shows the cost of drawing tokens at random instead of taking the most likely one. The JAX measurements generate device random bits with a single operation instead of JAX's default Threefry generator [38], a choice that materially affects sampling cost [37]. The cost is proportionally larger on four chips because the sampler does not speed up with more chips while the model does. The PyTorch server pays less for the `top_p` and `top_k` filters than the JAX server.

Libtt also supports speculative decoding, and Table 12 shows the gain. Under SGLang-JAX on four chips, the model's own draft layer speeds up a short factual prompt by 2.6× and a long summary by 24%. The gain depends on how often the draft is right.

## 11. Framework Compatibility

Libtt is designed to run existing JAX and PyTorch programs, and the applications built on them, without source changes. We test this against upstream framework test suites, which encode the frameworks' own definition of correct behavior, and against complete training and serving workloads. The framework interfaces are described in Section 8; the results here measure how much existing software they run.

Table 13 is a snapshot of suite coverage on one chip at the time of writing. Exclusions and test adaptations are listed below. The counts cover the selected suites and versions, not every feature of either framework.

**Table 12.** Speculative decoding, one request, tokens per second without and with the draft. MTP uses the model's own prediction layer as the draft (five draft tokens a step under SGLang-JAX, three under SGLang-TorchTPU); DFlash uses a separate draft model.

| | | SGLang-JAX | | SGLang-TorchTPU | |
|---|---|---|---|---|---|
| Request | Chips | without | with | without | with |
| *MTP, Qwen3.8-27B* | | | | | |
| 5-token prompt, greedy | 4 | 46.0 | 119.8 | 42.0 | 82.3 |
| 19-token code request, greedy | 4 | 46.0 | 102.7 | 42.0 | 82.2 |
| 215-token summary request, greedy | 4 | 45.7 | 56.8 | 41.8 | 72.7 |
| 5-token prompt, temperature 0.7 | 4 | 42.8 | 100.7 | 39.6 | 61.6 |
| 5-token prompt, greedy | 2 | 26.7 | 72.2 | 25.3 | 50.2 |
| 215-token summary request, greedy | 2 | 26.6 | 38.6 | 25.3 | 46.5 |
| *DFlash, Qwen3-8B* | | | | | |
| 19-token code request, greedy | 1 | 44.6 | 114.7 | 44.3 | 95.2 |
| 215-token summary request, greedy | 1 | 43.8 | 58.3 | 43.6 | 57.0 |

**Table 13.** Framework test suites on one chip, November 22, 2026: libtt 0.2.2 against upstream tt-xla 1.5.0.dev20260831. Under JAX 0.11.2 and `torch_tpu` 0.2.0, tt-xla crashes when it creates the client. Tests that crashed the process are counted as failed. PyTorch coverage is measured with the `torch_tpu` suite.

| Suite | Stack | Passed | Failed | Errors | Skipped |
|---|---|---|---|---|---|
| JAX 0.11.2 | libtt | 26,110 | 1,663 | 0 | 6,737 |
| | tt-xla | crashes at start-up | | | |
| JAX 0.7.1 | libtt | 24,075 | 1,577 | 0 | 6,511 |
| | tt-xla | 17,824 | 7,816 | 0 | 6,523 |
| `torch_tpu` 0.2.0 | libtt | 7,324 | 1,677 | 14 | 243 |
| | tt-xla | crashes at start-up | | | |

### 11.1. JAX

Supported programs use the ordinary JAX interface, including `jax.jit`, with libtt selected as the backend. To check this interface, the build fetches the source of the JAX version in use and runs its tests against the plugin without changing the test files. Thirteen files or directories that require facilities outside the plugin, like Pallas or CUDA, are excluded.

The suite exercises the behavior that existing programs depend on, including indexing, reductions, sorting, and control flow. It has been the main way of finding missing operations and incorrect semantics in the compiler. The changes in Section 5 broaden the range of programs that compile and produce the expected results.

### 11.2. PyTorch through TorchTPU

Libtt supports PyTorch eager execution and `torch.compile` through TorchTPU (Section 8.2). The compatibility measurements use that backend's test suite. Each file runs in a separate process, and operations are compared with CPU results using three samples for each operation and type. The original tests use reference files recorded on other hardware; two small patches adapt them for these comparisons, and four files are excluded.

The one-chip run includes failures from tests that require dynamic shapes or TPU-specific features. Its totals therefore reflect both unsupported behavior and differences between the test environment and the hardware the suite targets. Fixing more of these tests is an ongoing process.

### 11.3. SGLang and complete applications

The two SGLang servers provide application-level compatibility checks. SGLang-JAX uses a Tenstorrent backend that calls libtt's kernels. The supported SGLang-TorchTPU paths run without source edits: Libtt supplies adapters for TPU-specific operations. Section 10 describes these adaptations and the supported model families; Section 10.5 reports their performance.

Complete-model checks also exercise numerical behavior. Qwen3-8B scored 75.10 percent on MMLU, answering 10,331 of 13,757 questions correctly [39]. Training comparisons use the same saved starting point on chip and CPU, with smaller models and selected layers (Section 9.3). The TorchTitan fine-tuning run in Section 9.4 exercises checkpoint loading, training, export, serving, and evaluation. Together these checks provide evidence that the supported applications work through the full stack; they do not establish compatibility with every model or configuration.

## 12. Conclusion

Libtt brings existing JAX and PyTorch programs to Tenstorrent through a shared compiler and runtime so that the software that people already run on JAX and PyTorch can run on Tenstorrent hardware without being rewritten. Libtt extends Tenstorrent's stack with framework compatibility, numerical support, and execution across chips, and with the adaptations that existing training and serving systems need. The results show that this holds for complete applications: TorchTitan fine-tuning, SGLang serving, and Miles reinforcement learning, from loading weights and training to serving and evaluation.

The approach works because libtt reuses existing interfaces and does not ask applications to adopt new ones. The TorchTPU backend supplies PyTorch's eager and compiled execution models, and libtt replaces the TPU-specific operations that supported applications depend on, so each new application needs only adapters. These adapters have to be kept in step with the software they support, and libtt checks them against upstream framework tests, which expose assumptions that chip-specific tests miss.

### 12.1. Looking ahead

The results point to several directions that would widen what libtt can run, how efficiently it runs, and the scale it reaches.

*Framework coverage and numerics.* Operation coverage would extend to eigenvalue decompositions, more complex-number arithmetic, and the remaining windowed reductions. Dynamic shapes could be added, though most JAX and TorchTPU workloads already run with fixed shapes. Numerical work includes true 64-bit types (currently 32-bit except in random-number-generator state), BF16 operations that turn NaNs into infinities, FP8 subnormals, and the unsupported `float8_e8m0fnu` format.

*Memory use and execution.* Extending buffer reuse could reduce memory requirements further. Eligible donated inputs can already be released after their last use in single-device execution, but results still receive separate allocations, and inputs retained elsewhere and multi-device execution need further work. Moving loops whose termination depends on computed values onto the device would also remove the host from the critical path, since it currently drives them one iteration at a time.

*Distributed execution and hardware coverage.* Execution across machines would support deployments larger than the chips in one host. Related work includes automatic link recovery after a server is interrupted and support for processors beyond Blackhole.

*Training validation and scale.* The CPU comparisons cover a few steps on smaller models and selected layers and a scaled-down fine-tuning run (Section 9, Section 9.4). Reference runs of full-size training on other hardware, and longer runs across more models and configurations, would establish convergence and scaling. Larger models also need further memory reductions or additional parallelism, such as context parallelism.

*Supporting a kernel language.* Currently Pallas is the most widely used language to write kernels for TPUs. We could try to adapt it to produce kernels for Tenstorrent accelerators, or port another kernel language like TileLang.

*Improving fusion support.* Most fusions today are implemented by pattern-matching a subgraph and lowering it to a fixed, hand-written fused kernel. This is brittle: each variation in how a graph is written, such as a slightly different op sequence or a different data type, requires extending both the matcher and the kernel. Upstream already provides D2M, a mechanism that generates fused kernels automatically, without templates. It is currently disabled in libtt for three reasons: it significantly slows compilation, the code it generates is not yet fast enough, and it does not yet cover most of our existing fusions. We plan to adapt D2M so that it can replace most of our custom fusions without regressing compile time or kernel performance.

*Generalizing the compiler to other accelerators.* We believe the mechanisms in the compiler are broadly applicable to executing PyTorch and JAX programs on any tile-based accelerator. Writing a backend for TPUs could lead to an open-source drop-in replacement for `libtpu`. It might even make sense to write backends for AMD and NVIDIA GPUs and keep the tensors tiled in GPU memory: Tensors could directly be streamed into tensor cores, and furthermore it would lead to a common software stack for tile-based accelerators and GPUs.

Contributing the generally useful compiler and runtime changes upstream would make them available to every Tenstorrent user and reduce the cost of maintaining libtt's patches.